\documentclass[fleqn,10pt]{wlscirep}
\usepackage[utf8]{inputenc}
\usepackage[T1]{fontenc}
\usepackage{svg} 
\title{Distributed network of optically pumped magnetometers for space weather monitoring}

\author[1,*]{M. S. Mrozowski}
\author[1]{A. S. Bell}
\author[1]{P. F. Griffin}
\author[1]{D. Hunter}
\author[2]{D. Burt}
\author[1]{J. P. McGilligan}
\author[1]{E. Riis}
\author[3]{C. Beggan}
\author[1]{S. J. Ingleby}

\affil[1]{Department of Physics, SUPA, University of Strathclyde, Glasgow G4 0NG, United Kingdom}
\affil[2]{Kelvin Nanotechnology, 70 Oakfield Avenue, Glasgow G12 8LS, United Kingdom}
\affil[3]{British Geological Survey, The Lyell Centre, Research Avenue South, Edinburgh EH14 4BA, United Kingdom}

\affil[*]{marcin.mrozowski@strath.ac.uk}

\begin{abstract}
Spatial variation in the intensity of magnetospheric and ionospheric fluctuation during solar storms creates ground-induced currents, of importance in both infrastructure engineering and geophysical science. This activity is currently measured using a network of ground-based magnetometers, typically consisting of extensive installations at established observatory sites. We show that this network can be enhanced by the addition of remote quantum magnetometers which combine high sensitivity with intrinsic calibration. These nodes utilize scalable hardware and run independently of wired communication and power networks. We demonstrate that optically pumped magnetometers, utilizing mass-produced and miniaturized components, offer a single scalable sensor with the sensitivity and stability required for space weather observation. We describe the development and deployment of an off-grid magnetic sensing node, powered by a solar panel, present observed data from periods of low and high geomagnetic activity, and compare it to existing geomagnetic observatories.
\end{abstract}
\begin{document}

\flushbottom
\maketitle
%
%
\thispagestyle{empty}

\section*{Introduction}

Solar activity, resulting in the ejection of charged particles, induces a range of terrestrial effects, including geomagnetic storms \cite{REAENG2023}, ionospheric transients \cite{Kutiev2013,Bergeot2013} and ground-induced currents \cite{Pirjola2005}. These effects of space weather can, in turn, disrupt vital infrastructure, such as satellite communication and location systems \cite{Pajares2011,Cerruti2008}, terrestrial power distribution \cite{Oughton2019,Kelly2017,Erinmez2002}, telecommunication networks \cite{Medford1989}, and transport networks \cite{Wik2009,Eroshenko2010}. In particular, ground-induced potentials of only a few V/km could significantly increase railway signaling error rates \cite{Patterson2023}.

\begin{figure}
    \centering
    \includegraphics[width=0.7\linewidth]{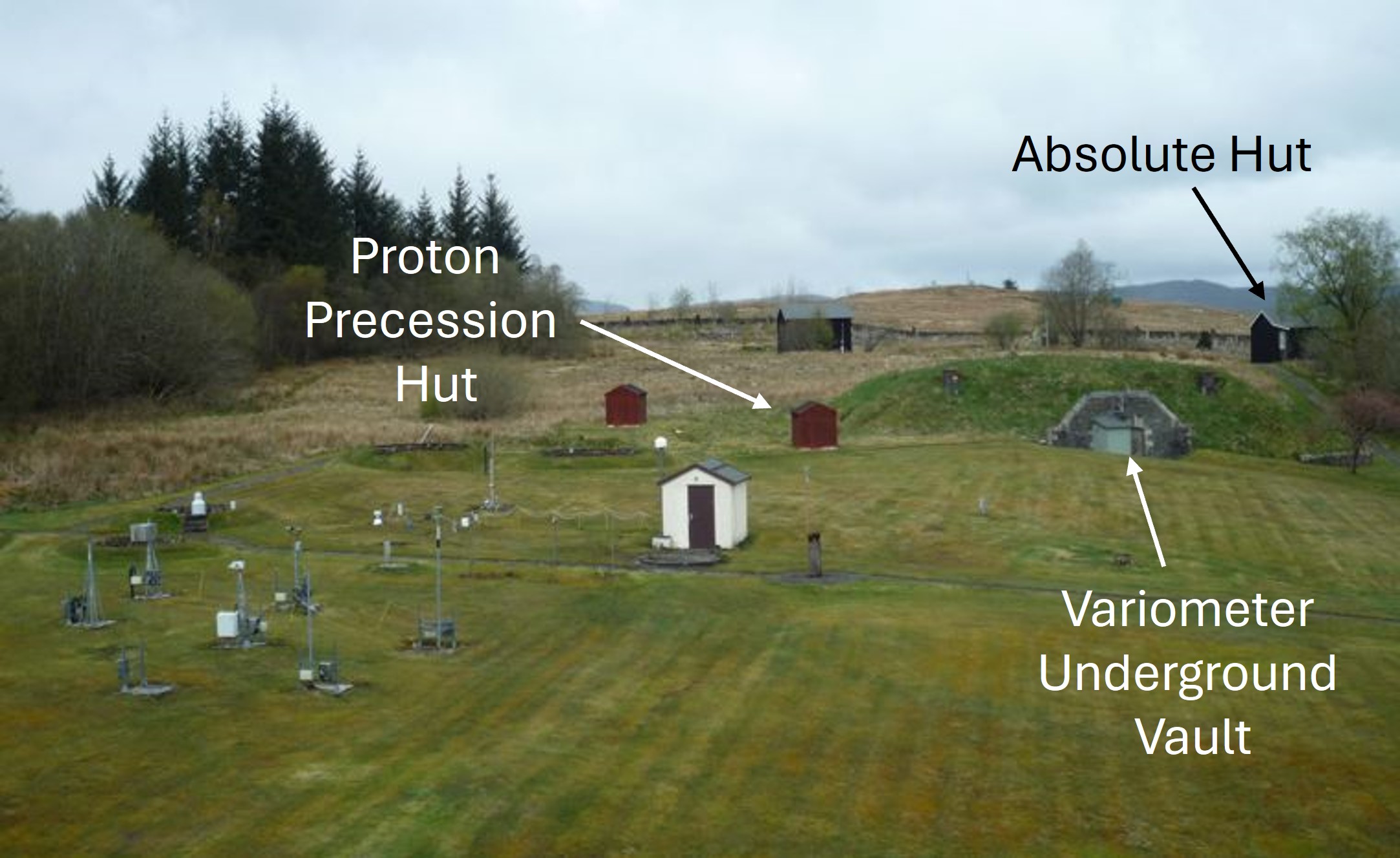}
    \caption{An annotated photograph showing the placement of different magnetometers on the Eskdalemuir BGS observatory including a proton precession magnetometer, a three-axis variometer, and a single-axis fluxgate theodolite. The proton magnetometer and fluxgate theodolite have been placed in above ground huts while a tri-axial fluxgate variometer is located in an underground vault that is temperature stabilized \cite{GDAS2002}. The site also hosts the UK Meteorological Office weather instruments in the left foreground.}
    \label{fig:ESK_site}
\end{figure}

The underlying geomagnetic variations are monitored in real-time by an international observatory network (INTERMAGNET) \cite{INTERMAGNET}. A typical geomagnetic observatory includes a range of magnetometers, combined to provide a stable, well-calibrated, sensitive record of the local geomagnetic field \cite{GDAS2002}. Three-axis vector fluxgate magnetometers are used in conjunction with inductive search coils, allowing higher bandwidth measurements, which, enable observation and monitoring of global lightning storms, manifested as Schumann resonance, which is typically represented as a fundamental peak at 7.86~Hz and its harmonics \cite{Fullekrug1995}. These sensors are complemented by scalar Overhauser proton magnetometers for increased stability \cite{Overhauser1953,Abragam1955}, establishing a calibrated baseline. This combination of sensors, installed in a well-controlled magnetic and thermal environment, enables geomagnetic measurement with high accuracy and precision. However, Overhauser proton magnetometers have limited bandwidth, while fluxgate magnetometers are not absolute, requiring weekly calibration \cite{Love2008}. Figure.~\ref{fig:ESK_site} shows the configuration of geomagnetic sensors at the British Geological Survey (BGS) Eskdalemuir Observatory site.

From Fig.~\ref{fig:ESK_site}, it can be seen that a standard geomagnetic observatory consists of a number of huts, which house different sensors while providing protective, and temperature-stabilized environments to reduce unwanted instrumental effects, which can appear to be a natural variation of the field. This approach increases the size, complexity, and cost of the observatory site, limiting the number of observatories that may be established globally.

Optically pumped magnetometers (OPMs), based on the detection of magneto-optical resonance in optically pumped atomic vapors, offer a combination of high sensitivity and low-drift measurements \cite{Ingleby2022,Hunter2023,Griffith2010}. OPMs may also be realized using mass-produced microfabricated components, packaged in compact and portable devices \cite{Liew2004,Kitching2008}. The potential for combining the functionality of a geomagnetic observatory site into a single instrument exists, and here we demonstrate OPMs as distributed nodes, independent of mains power or observatory infrastructure, added to the network of established geomagnetic observatories. At present, there are relatively few magnetic observatories globally (less than 200\cite{Love2013}) so the addition of new nodes, increasing the spatial resolution of geomagnetic field measurements, will have practical benefits in understanding the relationships between geomagnetic transients and ground-induced current flows. This is particularly true at mid- to high-latitudes where sub-100 km structures appear in the auroral oval as it moves equatorwards during strong geomagnetic activity \cite{Smith2019}. Finer scale resolution allows the most destructive features, such as highly curved magnetic transients, to be detected in real-time, which can provide rapid and localized warnings to infrastructure operators \cite{Dimmock2020}.

\section*{Results}
\subsection*{Space weather observation}\label{subsec:space_weather}
The off-grid magnetometer was set up at Finlaggan on the Isle of Islay in Scotland (55.8\textdegree~N, 6.2\textdegree~W). This site was selected as it is remote from 50~Hz power line and additional artificial magnetic noise contributions, providing an optimal background for observation of the effects of space weather. The control module and the sensor head described earlier were set up and fixed to a wooden platform. The platform with the sensor was orientated along the horizontal component of the Earth's magnetic field and fixed to the ground with wooden stakes to prevent any further movement. The sensor was inclined at $\approx$~70\textdegree~from the vertical plane of the platform. The PV panel was directed to face South and tilted at an angle of $\approx$~35\textdegree~from the vertical plane of the ground. A steeper panel angle would increase the sunlight capture efficiency. However, this angle was selected as it reduces the surface area of the panel along the vertical direction, making it less susceptible to damage from strong winds that can affect this location in winter. 
A wooden fence was erected around the site to prevent wildlife from disturbing the experiment.

Magnetic field measurements were sampled from the OPM at a sampling frequency of 50~S/s. A file was created every three hours totaling 540~kS after which it was filtered and downsampled to 1~S/s, to match the cadence of the BGS measurements. Both the 50~S/s data and 1~S/s data were saved to the local storage of the PC. Downsampled data were then uploaded to a cloud server as a timestamped file following Coordinated Universal Time (UTC) from the internal PC clock, synchronized daily to time from the internet. The downsampling was performed to reduce network traffic, as only the 1~S/s data files were uploaded to the server.

\begin{figure}
    \centering
    \includegraphics[width=0.5\linewidth]{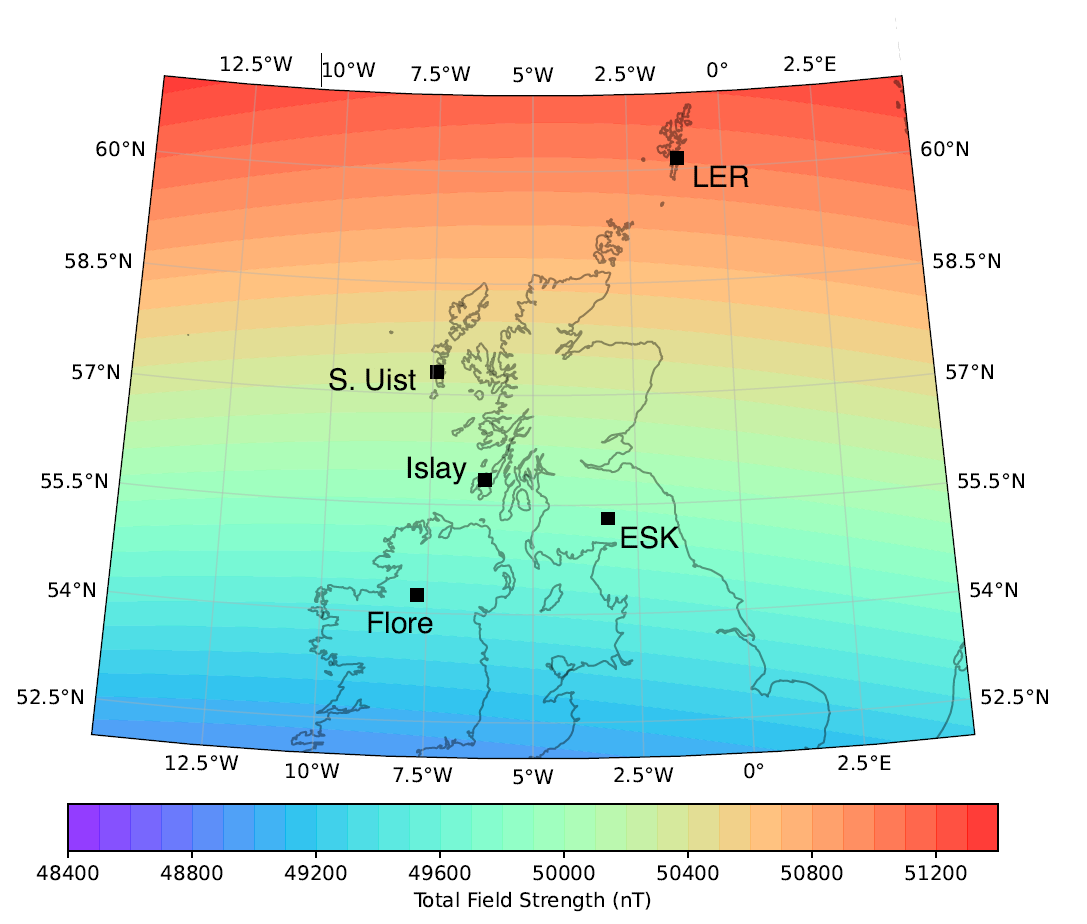}
    \caption{Map showing the location of the magnetometer network where observations were made. The colored contours show the expected background geomagnetic field magnitude for 01-Jan-2024, calculated using the CHAOS-7 geomagnetic field model \cite{Finlay2020}.}
    \label{fig:map}
\end{figure}

\begin{figure}[ht]
    \centering
    \includegraphics[scale = 0.584]{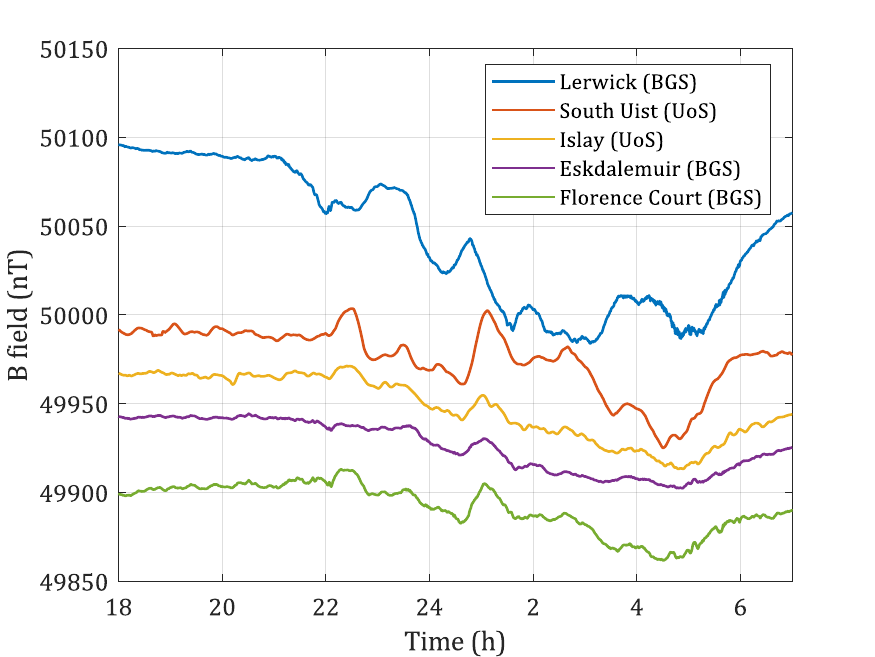}
    \caption{Magnetic data taken from 18:00 04-Oct-2023 to 7:00 05-Oct-2023 (UTC) at different sites. Lerwick, Eskdalemuir and Florence Court data were provided by the BGS. Data are ordered by latitude offset from one another with respect to Eskdalemuir data to improve readability. This data set shows "elevated activity" space weather conditions.}
    
    \label{fig:space_weather_active}

\end{figure}

The off-grid OPM setup on Islay was operational from 16:00 (UTC) on 04-Oct-2023. At the same time, a similarly configured OPM was in operation on the South Uist in the UK (57.4\textdegree~N, 7.4\textdegree~W). The test site was located closer to the mains power grid infrastructure compared to the Islay setup.
Elevated magnetic activity was detected at around 22:30 on 04-Oct-2023 which lasted until 4:00 on 05-Oct-2023, with a maximum Kp-index of 5 observed \cite{Matzka2021}. Kp-index is a measure of global geomagnetic activity derived from 3-hour measurements from ground-based geomagnetic observatories around the world. It ranges from 0 to 9, where 0 means very low geomagnetic activity and 9, extreme geomagnetic storms.
Variations in the magnitude of the geomagnetic field during that time interval were measured simultaneously by the Islay and South Uist OPMs as well as BGS observatories at Lerwick, Eskdalemuir, and a tri-axial variometer at Florence Court.
The map showing the location of the magnetometer network, and expected background geomagnetic fields is presented in Fig.~\ref{fig:map} and data from these locations are shown in Fig.~\ref{fig:space_weather_active}.

The main magnetic source of external field variation in the northern UK is the auroral electrojet, particularly during magnetic storms \cite{Beggan2015}. Due to the varying distance of the sensors from the auroral oval in this region, each records a different response - generally larger the closer its location to the auroral oval is. In addition, induced magnetic fields in the local subsurface contribute to smoothing or band-passing the magnetic field measured, as it is well known in Eskdalemuir \cite{Hutton1977}.

\begin{figure}[ht]
    \centering
    \includegraphics[scale = 0.584]{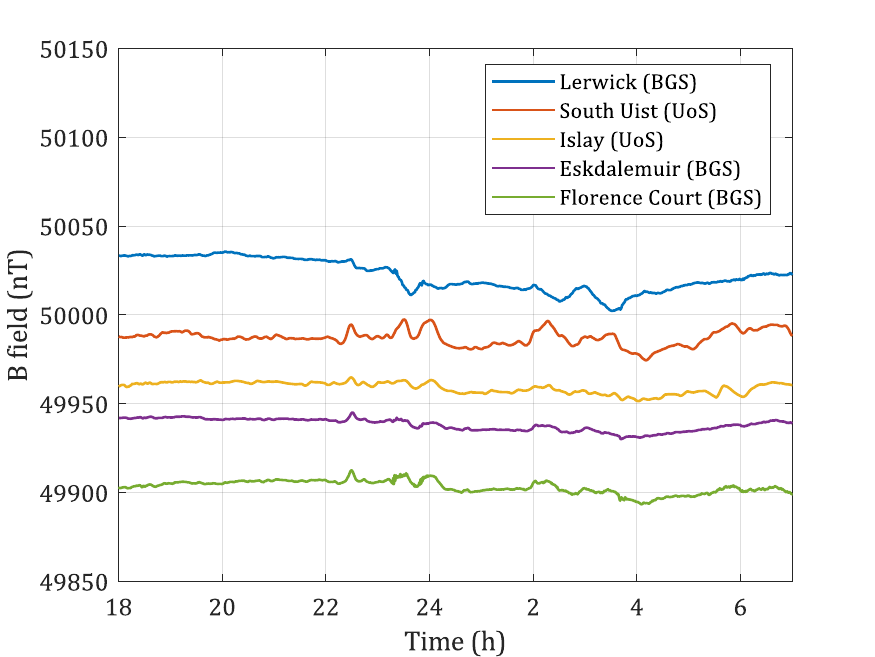}
    \caption{Magnetic data taken from 18:00 05-Oct-2023 to 7:00 06-Oct-2023 (UTC) at different sites. Lerwick, Eskdalemuir and Florence Court data were provided by the BGS. Data are ordered by latitude and offset from one another with respect to Eskdalemuir data to improve readability. This data set shows "low activity" space weather conditions.}
    
    \label{fig:space_weather_quiet}
    
\end{figure}

To present a comparison between an elevated magnetic activity as seen in Fig.~\ref{fig:space_weather_active} and typical magnetic activity, another set of data was taken the next day, over the same time frame. The resulting data are presented in Fig.~\ref{fig:space_weather_quiet}.
Here, it can be seen that magnetic deviation is much smaller ($\leq$~10~nT) in comparison to the magnetically active period as seen in Fig.~\ref{fig:space_weather_active} where deviation exceeds $\geq$~50~nT, from the quiet time background.

From Fig.~\ref{fig:space_weather_active}, it can be seen that all the sensors differ in response, which is attributed to the physical location of test sites, however, some common features exist. The diversity of data obtained from different sites highlights the importance of high spatial resolution for space weather monitoring.

The variation of the magnetic field over~100~km scales means that local enhancements of the field can be accurately measured. This feeds into better models of the induced geoelectric field, which poses a threat to the grounded infrastructure \cite{Hubert2024}.

\section*{Discussion}

The paper described the creation of a first-known field-deployed optically pumped magnetometer system for monitoring space weather. The Islay device was set up in a remote location, far from man-made magnetic noise sources (power lines, vehicles, etc.) while South Uist was closer to mains infrastructure. The deployed system was able to record space weather activity which was later compared to the data gathered from existing BGS observatories/variometers showing similar trends and also differences between observatories, associated with the subsurface geology and geographical location of the devices.
\\ 
Due to the simplicity of the system architecture, further identical OPM systems could be deployed in the future, creating a network of observatories providing better coverage of the geomagnetic events. The scalability of the approach was demonstrated by deploying another OPM system on South Uist which has recorded some events not seen by other observatories due to differences in their physical location, highlighting the importance of increasing spatial resolution through a wide network of observatories. 
\\
The system could benefit from additional improvements in the area of signal processing. By moving to a dedicated field-programmable gate array (FPGA) system, the level of integration would be increased, allowing for the replacement of the DAQ and PC in setup. This in turn would allow for lower power consumption of the system. The PC and DAQ currently used, limit the effective cadence of the sensor to $\approx$~140~Hz. By moving to a dedicated hardware processing system, higher sensor performance in terms of bandwidth could be achieved.\\

\section*{Methods}

\subsection*{Optically pumped magnetometer with a micromachined (MEMS) cell}\label{subsec:OPM}

\begin{figure}[ht]
    \centering
    \includegraphics[scale = 1.1]{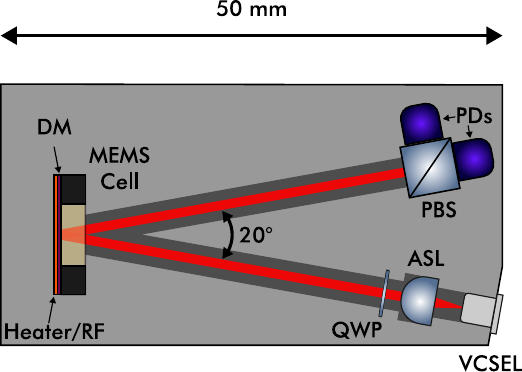}
    \caption{Simplified drawing of the OPM sensor head including vertical-cavity-surface-emitting-laser \textbf{(VCSEL)}, aspheric lens \textbf{(ASL)}, quarter-waveplate \textbf{(QWP)}, micromachined cesium vapor cell \textbf{(MEMS cell)}, dielectric mirror \textbf{(DM)}, planar non-inductive ohmic heater/radio frequency field source \textbf{(Heater/RF)}, polarizing beam splitter \textbf{(PBS)} and a pair of photodiodes \textbf{(PDs)}. The angle between the VCSEL beam incidence and the resulting reflection (20\textdegree), as well as the total length of the package (50~mm) are indicated.}
    
    \label{fig:OPM}

\end{figure}

A simplified schematic of the physics package is shown in Fig.~\ref{fig:OPM}. The magnetometer used in this study is a radio-frequency, $\mathrm{M_X}$ configuration, compact OPM. The sensor uses a commercially available, single-mode vertical-cavity-surface-emitting-laser (VCSEL) diode, with an operating wavelength of 894.6~nm used to interrogate the $D_1$ line of the atomic vapor. The VCSEL is orientated at an angle of 10\textdegree~from the normal of the mirror attached to the back of the vapor cell and its resulting beam is collimated to 3.2~mm with an aspherical lens. The light is circularly polarized with a quarter-wave plate placed after the collimation stage.
The sensor uses a micro-machined vapor cell \cite{Dyer2022}, consisting of a glass-silicon-glass stack. The silicon wafer is machined to produce cavity dimensions of 6~mm$\times$6~mm$\times$3~mm, anodically bonded to borosilicate glass on each side for a hermetic seal. An azide solution is deposited into the cell, which is later dissociated by UV light, to realize a saturated vapor pressure of $^{133}$Cs and $\approx$~200~Torr of $\mathrm{N_2}$ which acts as a quenching/buffer gas. 
The beam passes through the vapor cell and is reflected by a dielectric mirror, which is adhered to the rear cell window with optical adhesive, enabling a double-pass configuration, increasing optical path length through the alkali vapor by a factor of two. This configuration also allows for the reduction of cell temperature and thus power consumption. Another benefit of double-pass configuration is the ability to situate the photodiode far away from the actual sensor, reducing stray magnetic noise pickup.  The beam exits the cell at a total reflected angle of 20\textdegree~from the VCSEL. The 20\textdegree~reflection angle was selected to minimize the package's overall size and reduce the number of components needed to realize the OPM. The cell is heated with a planar, non-inductively wound ohmic heater to 80~\textdegree$\mathrm{C}$ and is driven at an RF resonant angular frequency ($\omega_0 = \gamma \times B_0$, where $\gamma$~$\approx$~3.5~Hz/nT). The heater is non-inductively wound, however, at operational currents used for heating the cell, it induces a weak magnetic field along the beam propagation axis. The resultant field acts as a source of RF resonant magnetic field for the magnetometer. The reflected beam is then incident onto the polarizing beam splitter where its resulting polarization rotation is detected on a differential photodiode.  

The optics package is made out of 3D-printed material housed in an additional case to provide mounting points and protection from adverse weather conditions. 3D-printing is often used for manufacturing OPM sensor heads due to precision, mechanical robustness and inherent non-magnetic properties of the thermoplastics/resins used \cite{Carter2016,Dawson2023}. The interface to the sensor is provided with two, $\approx$~1~m-long RJ45 cables. The sensor head is an evolution of the previously published design \cite{Ingleby2022}.

\begin{figure}[ht]
	\includegraphics[width=1\linewidth]{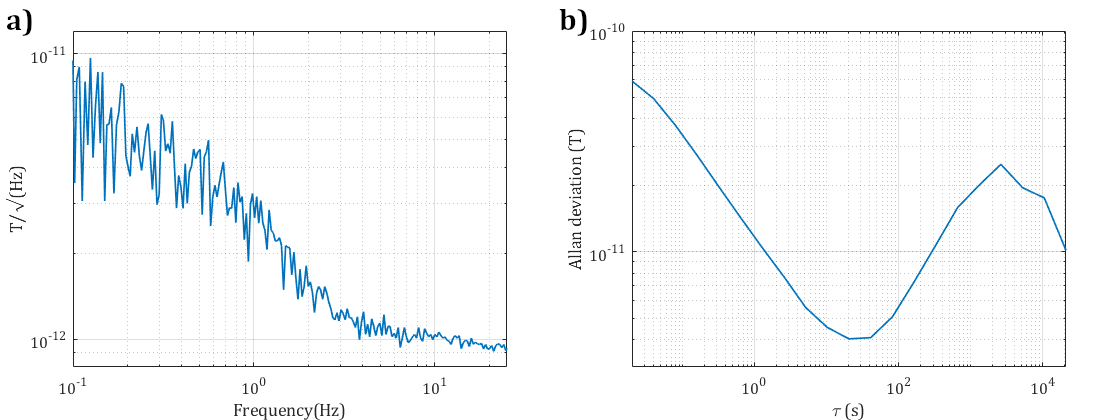}
	\caption{Noise spectral density of the OPM in a bandwidth of 0.1~-~25~Hz in a) and Allan Deviation of the OPM running for 12 hours in b). The measurements were performed in a 5-layer mu-metal shield, with a 50~$\mu$T bias field present along the sensitive axis of the magnetometer. The applied field was derived from a set of Helmholtz coils present in the shield, driven with a custom low noise current driver \cite{Mrozowski2023}. The sensor exhibits a sensitivity of $\approx$~9~pT/$\sqrt{\mathrm{Hz}}$ at 0.1~Hz and $\approx$~3~pT/$\sqrt{\mathrm{Hz}}$ at 1~Hz. The sensitivity of the sensor meets the one-second INTERMAGNET standard\cite{INTERMAGNET_1s}. The sensor achieves the best stability after $\approx$~20~seconds of averaging, after which drift dominates.}
	\label{fig:OPM_performance}
\end{figure}

The sensor performance is presented in Fig.~\ref{fig:OPM_performance} showing sensitivity and its stability with an overlapping Allan deviation \cite{Allan1966}. The sensitivity of the sensor is within the limits of the one-second INTERMAGNET standard \cite{INTERMAGNET_1s} that the international network of magnetic observatories meets.

\subsection*{Field deployed optically pumped magnetometer setup}\label{sec:off_grid_setup}

\begin{figure}[ht]
\centering\includegraphics[scale=1.4]{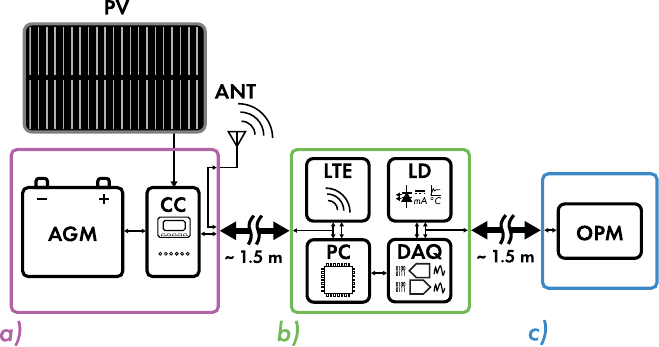}
\caption{Simplified block diagram of the off-grid OPM setup split into three main modules: (a) Power supply, (b) Control and processing, and (c) Sensor head. Each is housed in an IP67-rated enclosure and connected with fully plastic and non-magnetic IP67-rated conduits. The setup includes: Lead-acid battery \textbf{(AGM)}, charge controller \textbf{(CC)}, personal computer \textbf{(PC)}, 4G modem \textbf{(LTE)}, Data acquisition \textbf{(DAQ)}, laser driver \textbf{(LD)}, optically pumped magnetometer \textbf{(OPM)}. The setup is powered by a photovoltaic panel \textbf{(PV)} covering the power supply module. The antennas \textbf{(ANT)} used for LTE communication are present on the enclosure housing the power supply.}
\label{fig:System_setup}
\end{figure}

The simplified block diagram of the off-grid OPM setup is presented in Fig.~\ref{fig:System_setup}. The system can be broken into three modules: power supply, control and processing, and sensor head.

The power supply module houses a lead acid battery and a charge controller. The battery has a total capacity of 160~Ah (1920~Wh) and is of absorbent glass mat (AGM) type. The AGM type was selected as it features good charge retention under changing environmental conditions that the setup will be exposed to in the field. The battery is charged using a monocrystalline photovoltaic (PV) panel, capable of supplying a maximum power of 400~W. The monocrystalline architecture offers better efficiency than polycrystalline panels \cite{Wenham2013}. During testing, the panel was capable of providing between $\approx$~30~-~70~W in the daytime, during overcast weather, at a latitude 55.8\textdegree~N. Less power can be obtained from late December into early January when the daytime is shortest and temperatures are much lower. 
The battery is charged with a charge controller (EPEVER, Triron3210N) featuring maximum power point tracking (MPPT), enabling more efficient energy extraction from shaded PV panel. The controller also enables remote monitoring of the PV as well as the battery with a PC. The power output from the charge controller is connected to the control module through a tightly wound shielded twisted pair to reduce the magnetic contribution from the power supply. 
The battery and the charge controller are housed in an IP67-rated plastic enclosure, which is additionally covered by the PV panel to further protect it from atmospheric conditions.

The control and processing module houses a mini portable computer (Compulab, fitlet2), a data acquisition board (Digilent, Analog Discovery 2), a custom laser driver, and an LTE modem (Multitech, Multiconnect Cell 100) for remote logging of data. The control system is housed in a die-cast aluminum enclosure, improving electromagnetic interference (EMI) rejection and improving cooling of all the components inside. The enclosure is rated to IP67 standard.
The PC is powered directly from the charge controller as it accepts voltages in a range of 9~-~36~V which is within the operating range of AGM batteries. The PC has a typical power consumption of $\leq$~10~W. The PC is used to control all the peripherals, including the OPM. The DAQ and the laser driver are controlled through LabVIEW software. LTE antennas are wired out to the power module as it is housed in the tallest enclosure where they can be mounted externally, while at the same time, they can be covered from rain with the PV panel. The OPM as well as the control subsystem consume $\leq$~5~W of power, bringing the total average power consumption of the system to $\approx$~15~W.

The sensor head module houses the OPM described earlier. The enclosure is made out of polycarbonate, features an IP67 rating, and has been modified to be fully non-magnetic by replacing stainless steel bolts with nylon ones. As the enclosure is plastic, it provides better insulation to the OPM from changes in ambient temperature and improves its stability. Inside the enclosure, the sensor is mounted using a custom 3D printed mount which allows it to be tilted at angles between 0\textdegree~-~90\textdegree, enabling alignment of its sensitive axis with the Earth's magnetic field vector.

To provide communication and power between each module, cables were routed through non-magnetic, IP67-rated, 25~mm outer-diameter conduits. The power supply module and the processing module are separated by $\approx$~1.5~m-long conduit, while the sensor head is distanced by $\approx$~1.5~m-long from the control module. The separation provided between each module helps in isolating the OPM from the magnetic contribution present from magnetic components such as the battery, as well as DC powering the control module.

\section*{Data Availability Statement}

The datasets used in this work are available at (https://doi.org/10.15129/4c091e48-f698-42c0-9863-e0b52fcd21c8). BGS magnetic data is publicly available at (www.bgs.ac.uk).

\bibliography{space_weather}

\begin{thebibliography}{10}
\urlstyle{rm}
\expandafter\ifx\csname url\endcsname\relax
  \def\url#1{\texttt{#1}}\fi
\expandafter\ifx\csname urlprefix\endcsname\relax\def\urlprefix{URL }\fi
\expandafter\ifx\csname doiprefix\endcsname\relax\def\doiprefix{DOI: }\fi
\providecommand{\bibinfo}[2]{#2}
\providecommand{\eprint}[2][]{\url{#2}}

\bibitem{REAENG2023}
\bibinfo{title}{Extreme space weather: impacts on engineered systems and infrastructure}.
\newblock \bibinfo{type}{Tech. Rep.}, \bibinfo{institution}{Royal Academy of Engineering} (\bibinfo{year}{2023}).

\bibitem{Kutiev2013}
\bibinfo{author}{Kutiev, I.} \emph{et~al.}
\newblock \bibinfo{journal}{\bibinfo{title}{{Solar activity impact on the Earth’s upper atmosphere}}}.
\newblock {\emph{\JournalTitle{{Journal of Space Weather and Space Climate}}}} \textbf{\bibinfo{volume}{3}}, \bibinfo{pages}{A06}, \doiprefix\url{10.1051/SWSC/2013028} (\bibinfo{year}{2013}).

\bibitem{Bergeot2013}
\bibinfo{author}{Bergeot, N.} \emph{et~al.}
\newblock \bibinfo{journal}{\bibinfo{title}{The influence of space weather on ionospheric total electron content during the 23rd solar cycle}}.
\newblock {\emph{\JournalTitle{Journal of Space Weather and Space Climate}}} \textbf{\bibinfo{volume}{3}}, \bibinfo{pages}{A25}, \doiprefix\url{10.1051/SWSC/2013047} (\bibinfo{year}{2013}).

\bibitem{Pirjola2005}
\bibinfo{author}{Pirjola, R.}, \bibinfo{author}{Viljanen, A.}, \bibinfo{author}{Pulkkinen, A.}, \bibinfo{author}{Kilpua, S.} \& \bibinfo{author}{Amm, O.}
\newblock \bibinfo{title}{Ground effects of space weather}.
\newblock In \bibinfo{editor}{Daglis, I.~A.} (ed.) \emph{\bibinfo{booktitle}{Effects of Space Weather on Technology Infrastructure}}, \bibinfo{pages}{235--256} (\bibinfo{publisher}{Springer Netherlands}, \bibinfo{address}{Dordrecht}, \bibinfo{year}{2005}).

\bibitem{Pajares2011}
\bibinfo{author}{Hernández-Pajares, M.} \emph{et~al.}
\newblock \bibinfo{journal}{\bibinfo{title}{{The ionosphere: Effects, GPS modeling and the benefits for space geodetic techniques}}}.
\newblock {\emph{\JournalTitle{Journal of Geodesy}}} \textbf{\bibinfo{volume}{85}}, \bibinfo{pages}{887--907}, \doiprefix\url{10.1007/S00190-011-0508-5} (\bibinfo{year}{2011}).

\bibitem{Cerruti2008}
\bibinfo{author}{Cerruti, A.~P.} \emph{et~al.}
\newblock \bibinfo{journal}{\bibinfo{title}{{Effect of intense December 2006 solar radio bursts on GPS receivers}}}.
\newblock {\emph{\JournalTitle{Space Weather}}} \textbf{\bibinfo{volume}{6}}, \bibinfo{pages}{S10D07}, \doiprefix\url{10.1029/2007SW000375} (\bibinfo{year}{2008}).

\bibitem{Oughton2019}
\bibinfo{author}{Oughton, E.~J.} \emph{et~al.}
\newblock \bibinfo{journal}{\bibinfo{title}{{A Risk Assessment Framework for the Socioeconomic Impacts of Electricity Transmission Infrastructure Failure Due to Space Weather: An Application to the United Kingdom}}}.
\newblock {\emph{\JournalTitle{Risk Analysis}}} \textbf{\bibinfo{volume}{39}}, \bibinfo{pages}{1022--1043}, \doiprefix\url{10.1111/RISA.13229} (\bibinfo{year}{2019}).

\bibitem{Kelly2017}
\bibinfo{author}{Kelly, G.~S.}, \bibinfo{author}{Viljanen, A.}, \bibinfo{author}{Beggan, C.~D.} \& \bibinfo{author}{Thomson, A.~W.}
\newblock \bibinfo{journal}{\bibinfo{title}{{Understanding GIC in the UK and French high-voltage transmission systems during severe magnetic storms}}}.
\newblock {\emph{\JournalTitle{Space Weather}}} \textbf{\bibinfo{volume}{15}}, \bibinfo{pages}{99--114}, \doiprefix\url{10.1002/2016SW001469} (\bibinfo{year}{2017}).

\bibitem{Erinmez2002}
\bibinfo{author}{Erinmez, I.~A.}, \bibinfo{author}{Kappenman, J.~G.} \& \bibinfo{author}{Radasky, W.~A.}
\newblock \bibinfo{journal}{\bibinfo{title}{{Management of the geomagnetically induced current risks on the national grid company's electric power transmission system}}}.
\newblock {\emph{\JournalTitle{Journal of Atmospheric and Solar-Terrestrial Physics}}} \textbf{\bibinfo{volume}{64}}, \bibinfo{pages}{743--756}, \doiprefix\url{10.1016/S1364-6826(02)00036-6} (\bibinfo{year}{2002}).

\bibitem{Medford1989}
\bibinfo{author}{Medford, L.~V.}, \bibinfo{author}{Lanzerotti, L.~J.}, \bibinfo{author}{Kraus, J.~S.} \& \bibinfo{author}{Maclennan, C.~G.}
\newblock \bibinfo{journal}{\bibinfo{title}{{Transatlantic Earth Potential Variations During the March 1989 Magnetic Storms}}}.
\newblock {\emph{\JournalTitle{Geophysical Research Letters}}} \textbf{\bibinfo{volume}{16}}, \bibinfo{pages}{1145--1148}, \doiprefix\url{10.1029/GL016I010P01145} (\bibinfo{year}{1989}).

\bibitem{Wik2009}
\bibinfo{author}{Wik, M.} \emph{et~al.}
\newblock \bibinfo{journal}{\bibinfo{title}{{Space weather events in July 1982 and October 2003 and the effects of geomagnetically induced currents on Swedish technical systems}}}.
\newblock {\emph{\JournalTitle{Annales Geophysicae}}} \textbf{\bibinfo{volume}{27}}, \bibinfo{pages}{1775--1787}, \doiprefix\url{10.5194/ANGEO-27-1775-2009} (\bibinfo{year}{2009}).

\bibitem{Eroshenko2010}
\bibinfo{author}{Eroshenko, E.~A.} \emph{et~al.}
\newblock \bibinfo{journal}{\bibinfo{title}{{Effects of strong geomagnetic storms on Northern railways in Russia}}}.
\newblock {\emph{\JournalTitle{Advances in Space Research}}} \textbf{\bibinfo{volume}{46}}, \bibinfo{pages}{1102--1110}, \doiprefix\url{10.1016/J.ASR.2010.05.017} (\bibinfo{year}{2010}).

\bibitem{Patterson2023}
\bibinfo{author}{Patterson, C.~J.}, \bibinfo{author}{Wild, J.~A.} \& \bibinfo{author}{Boteler, D.~H.}
\newblock \bibinfo{journal}{\bibinfo{title}{{Modeling “Wrong Side” Failures Caused by Geomagnetically Induced Currents in Electrified Railway Signaling Systems in the UK}}}.
\newblock {\emph{\JournalTitle{Space Weather}}} \textbf{\bibinfo{volume}{21}}, \bibinfo{pages}{e2023SW003625}, \doiprefix\url{10.1029/2023SW003625} (\bibinfo{year}{2023}).

\bibitem{INTERMAGNET}
\bibinfo{title}{{INTERMAGNET International Real-Time Magnetic Observatory Network}}.
\newblock \bibinfo{howpublished}{\url{https://intermagnet.org/}} (\bibinfo{year}{2024}).
\newblock \bibinfo{note}{Accessed: 2024-02-06}.

\bibitem{GDAS2002}
\bibinfo{organization}{British Geological Survery}.
\newblock \emph{\bibinfo{title}{G-DAS Geomagnetic Data Acquisition System}} (\bibinfo{year}{2002}).

\bibitem{Fullekrug1995}
\bibinfo{author}{Füllekrug, M.}
\newblock \bibinfo{journal}{\bibinfo{title}{{Schumann resonances in magnetic field components}}}.
\newblock {\emph{\JournalTitle{{Journal of Atmospheric and Terrestrial Physics}}}} \textbf{\bibinfo{volume}{57}}, \bibinfo{pages}{479--484}, \doiprefix\url{10.1016/0021-9169(94)00075-Y} (\bibinfo{year}{1995}).

\bibitem{Overhauser1953}
\bibinfo{author}{Overhauser, A.~W.}
\newblock \bibinfo{journal}{\bibinfo{title}{Polarization of nuclei in metals}}.
\newblock {\emph{\JournalTitle{Physical Review}}} \textbf{\bibinfo{volume}{92}}, \bibinfo{pages}{411}, \doiprefix\url{10.1103/PhysRev.92.411} (\bibinfo{year}{1953}).

\bibitem{Abragam1955}
\bibinfo{author}{Abragam, A.}
\newblock \bibinfo{journal}{\bibinfo{title}{Overhauser effect in nonmetals}}.
\newblock {\emph{\JournalTitle{Physical Review}}} \textbf{\bibinfo{volume}{98}}, \bibinfo{pages}{1729}, \doiprefix\url{10.1103/PhysRev.98.1729} (\bibinfo{year}{1955}).

\bibitem{Love2008}
\bibinfo{author}{Love, J.~J.}
\newblock \bibinfo{journal}{\bibinfo{title}{Magnetic monitoring of earth and space}}.
\newblock {\emph{\JournalTitle{Physics Today}}} \textbf{\bibinfo{volume}{61}}, \bibinfo{pages}{31--37}, \doiprefix\url{10.1063/1.2883907} (\bibinfo{year}{2008}).

\bibitem{Ingleby2022}
\bibinfo{author}{Ingleby, S.}, \bibinfo{author}{Griffin, P.}, \bibinfo{author}{Dyer, T.}, \bibinfo{author}{Mrozowski, M.} \& \bibinfo{author}{Riis, E.}
\newblock \bibinfo{journal}{\bibinfo{title}{A digital alkali spin maser}}.
\newblock {\emph{\JournalTitle{Scientific Reports}}} \textbf{\bibinfo{volume}{12}}, \bibinfo{pages}{1--7}, \doiprefix\url{10.1038/s41598-022-16910-z} (\bibinfo{year}{2022}).

\bibitem{Hunter2023}
\bibinfo{author}{Hunter, D.} \emph{et~al.}
\newblock \bibinfo{journal}{\bibinfo{title}{Optical pumping enhancement of a free-induction-decay magnetometer}}.
\newblock {\emph{\JournalTitle{JOSA B}}} \textbf{\bibinfo{volume}{40}}, \bibinfo{pages}{2664--2673}, \doiprefix\url{10.1364/JOSAB.501086} (\bibinfo{year}{2023}).

\bibitem{Griffith2010}
\bibinfo{author}{Griffith, W.~C.} \emph{et~al.}
\newblock \bibinfo{journal}{\bibinfo{title}{Femtotesla atomic magnetometry in a microfabricated vapor cell}}.
\newblock {\emph{\JournalTitle{Optics Express}}} \textbf{\bibinfo{volume}{18}}, \bibinfo{pages}{27167--27172}, \doiprefix\url{10.1364/OE.18.027167} (\bibinfo{year}{2010}).

\bibitem{Liew2004}
\bibinfo{author}{Liew, L.~A.} \emph{et~al.}
\newblock \bibinfo{journal}{\bibinfo{title}{Microfabricated alkali atom vapor cells}}.
\newblock {\emph{\JournalTitle{Applied Physics Letters}}} \textbf{\bibinfo{volume}{84}}, \bibinfo{pages}{2694}, \doiprefix\url{10.1063/1.1691490} (\bibinfo{year}{2004}).

\bibitem{Kitching2008}
\bibinfo{author}{Kitching, J.} \emph{et~al.}
\newblock \bibinfo{journal}{\bibinfo{title}{Microfabricated atomic magnetometers and applications}}.
\newblock {\emph{\JournalTitle{2008 IEEE International Frequency Control Symposium, FCS}}} \bibinfo{pages}{789--794}, \doiprefix\url{10.1109/FREQ.2008.4623107} (\bibinfo{year}{2008}).

\bibitem{Love2013}
\bibinfo{author}{Love, J.~J.} \& \bibinfo{author}{Chulliat, A.}
\newblock \bibinfo{journal}{\bibinfo{title}{{An International Network of Magnetic Observatories}}}.
\newblock {\emph{\JournalTitle{Eos, Transactions American Geophysical Union}}} \textbf{\bibinfo{volume}{94}}, \bibinfo{pages}{373--374}, \doiprefix\url{10.1002/2013EO420001} (\bibinfo{year}{2013}).

\bibitem{Smith2019}
\bibinfo{author}{Smith, A.~W.}, \bibinfo{author}{Freeman, M.~P.}, \bibinfo{author}{Rae, I.~J.} \& \bibinfo{author}{Forsyth, C.}
\newblock \bibinfo{journal}{\bibinfo{title}{{The Influence of Sudden Commencements on the Rate of Change of the Surface Horizontal Magnetic Field in the United Kingdom}}}.
\newblock {\emph{\JournalTitle{{Space Weather}}}} \textbf{\bibinfo{volume}{17}}, \bibinfo{pages}{1605--1617}, \doiprefix\url{10.1029/2019SW002281} (\bibinfo{year}{2019}).

\bibitem{Dimmock2020}
\bibinfo{author}{Dimmock, A.~P.} \emph{et~al.}
\newblock \bibinfo{journal}{\bibinfo{title}{{On the Regional Variability of dB/dt and Its Significance to GIC}}}.
\newblock {\emph{\JournalTitle{Space Weather}}} \textbf{\bibinfo{volume}{18}}, \bibinfo{pages}{1–20}, \doiprefix\url{10.1029/2020SW002497} (\bibinfo{year}{2020}).

\bibitem{Matzka2021}
\bibinfo{author}{Matzka, J.}, \bibinfo{author}{Stolle, C.}, \bibinfo{author}{Yamazaki, Y.}, \bibinfo{author}{Bronkalla, O.} \& \bibinfo{author}{Morschhauser, A.}
\newblock \bibinfo{journal}{\bibinfo{title}{{The Geomagnetic Kp Index and Derived Indices of Geomagnetic Activity}}}.
\newblock {\emph{\JournalTitle{Space Weather}}} \textbf{\bibinfo{volume}{19}}, \bibinfo{pages}{1–21}, \doiprefix\url{10.1029/2020SW002641} (\bibinfo{year}{2021}).

\bibitem{Finlay2020}
\bibinfo{author}{Finlay, C.~C.} \emph{et~al.}
\newblock \bibinfo{journal}{\bibinfo{title}{{The CHAOS-7 geomagnetic field model and observed changes in the South Atlantic Anomaly}}}.
\newblock {\emph{\JournalTitle{Earth, Planets and Space}}} \textbf{\bibinfo{volume}{72}}, \bibinfo{pages}{1--31}, \doiprefix\url{10.1186/S40623-020-01252-9} (\bibinfo{year}{2020}).

\bibitem{Beggan2015}
\bibinfo{author}{Beggan, C.~D.}
\newblock \bibinfo{journal}{\bibinfo{title}{{Sensitivity of geomagnetically induced currents to varying auroral electrojet and conductivity models}}}.
\newblock {\emph{\JournalTitle{{Earth, Planets and Space}}}} \textbf{\bibinfo{volume}{67}}, \bibinfo{pages}{1--12}, \doiprefix\url{10.1186/S40623-014-0168-9} (\bibinfo{year}{2015}).

\bibitem{Hutton1977}
\bibinfo{author}{Hutton, V.~R.}, \bibinfo{author}{Sik, J.~M.} \& \bibinfo{author}{Gough, D.~I.}
\newblock \bibinfo{journal}{\bibinfo{title}{{Electrical conductivity and tectonics of Scotland}}}.
\newblock {\emph{\JournalTitle{Nature}}} \textbf{\bibinfo{volume}{266}}, \bibinfo{pages}{617--620}, \doiprefix\url{10.1038/266617a0} (\bibinfo{year}{1977}).

\bibitem{Hubert2024}
\bibinfo{author}{Hübert, J.} \emph{et~al.}
\newblock \bibinfo{journal}{\bibinfo{title}{Validating a uk geomagnetically induced current model using differential magnetometer measurements}}.
\newblock {\emph{\JournalTitle{Space Weather}}} \textbf{\bibinfo{volume}{22}}, \bibinfo{pages}{1--16}, \doiprefix\url{10.1029/2023SW003769} (\bibinfo{year}{2024}).

\bibitem{Dyer2022}
\bibinfo{author}{Dyer, S.} \emph{et~al.}
\newblock \bibinfo{journal}{\bibinfo{title}{Micro-machined deep silicon atomic vapor cells}}.
\newblock {\emph{\JournalTitle{Journal of Applied Physics}}} \textbf{\bibinfo{volume}{132}}, \bibinfo{pages}{134401}, \doiprefix\url{10.1063/5.0114762} (\bibinfo{year}{2022}).

\bibitem{Carter2016}
\bibinfo{author}{Carter, T.~R.} \emph{et~al.}
\newblock \bibinfo{journal}{\bibinfo{title}{Four-channel optically pumped atomic magnetometer for magnetoencephalography}}.
\newblock {\emph{\JournalTitle{Optics Express}}} \textbf{\bibinfo{volume}{24}}, \bibinfo{pages}{15403--15416}, \doiprefix\url{10.1364/OE.24.015403} (\bibinfo{year}{2016}).

\bibitem{Dawson2023}
\bibinfo{author}{Dawson, R.} \emph{et~al.}
\newblock \bibinfo{journal}{\bibinfo{title}{Portable single-beam cesium zero-field magnetometer for magnetocardiography}}.
\newblock {\emph{\JournalTitle{Journal of Optical Microsystems}}} \textbf{\bibinfo{volume}{3}}, \bibinfo{pages}{044501}, \doiprefix\url{10.1117/1.JOM.3.4.044501} (\bibinfo{year}{2023}).

\bibitem{Allan1966}
\bibinfo{author}{Allan, D.}
\newblock \bibinfo{journal}{\bibinfo{title}{Statistics of atomic frequency standards}}.
\newblock {\emph{\JournalTitle{Proceedings of the IEEE}}} \textbf{\bibinfo{volume}{54}}, \bibinfo{pages}{221--230}, \doiprefix\url{10.1109/PROC.1966.4634} (\bibinfo{year}{1966}).

\bibitem{INTERMAGNET_1s}
\bibinfo{title}{{INTERMAGNET International Real-Time Magnetic Observatory Network - Definitive 1-second standard}}.
\newblock \bibinfo{howpublished}{\url{https://intermagnet.org/docs/technical/im_tn_06_v1_0.pdf}} (\bibinfo{year}{2024}).
\newblock \bibinfo{note}{Accessed: 2024-02-06}.

\bibitem{Mrozowski2023}
\bibinfo{author}{Mrozowski, M.~S.}, \bibinfo{author}{Chalmers, I.~C.}, \bibinfo{author}{Ingleby, S.~J.}, \bibinfo{author}{Griffin, P.~F.} \& \bibinfo{author}{Riis, E.}
\newblock \bibinfo{journal}{\bibinfo{title}{Ultra-low noise, bi-polar, programmable current sources}}.
\newblock {\emph{\JournalTitle{Review of Scientific Instruments}}} \textbf{\bibinfo{volume}{94}}, \doiprefix\url{10.1063/5.0114760} (\bibinfo{year}{2023}).

\bibitem{Wenham2013}
\bibinfo{author}{Wenham, S.~R.}, \bibinfo{author}{Green, M.~A.}, \bibinfo{author}{Watt, M.~E.}, \bibinfo{author}{Corkish, R.~P.} \& \bibinfo{author}{Sproul, A.~B.}
\newblock \emph{\bibinfo{title}{Applied photovoltaics, third edition}} (\bibinfo{publisher}{Taylor and Francis}, \bibinfo{year}{2013}).

\end{thebibliography}

\section*{Acknowledgements}

The results presented in this paper rely on the data collected at Lerwick, Eskdalemuir and Florence Court. We thank the British Geological Survey for supporting its operation and INTERMAGNET for promoting high standards of magnetic observatory practice (www.intermagnet.org). This work was partially funded by EPSRC (grant numbers EP/X036391/1 and EP/T001046/1). J. P. M. gratefully acknowledges funding from a Royal Academy of Engineering Research Fellowship.

\section*{Author contributions statement}

M. M. designed the system, installed the field trial and analyzed the data, with input from S. J. I, D. H. and E. R. S. J. I. designed the magnetometer sensor head and wrote the sensor software. A. S. B. designed and built the field station hardware and located the site. D. B. and J. P. M. developed and fabricated the Cs vapor cell. C. B. added interpretation of the data and comparison with reference measurements. All authors discussed the results and commented on the manuscript.

\section*{Competing interests}
The authors declare no competing interests.

\end{document}